\documentclass{article}

\usepackage{latexsym}
\usepackage{amssymb}
\usepackage{amsmath}

\def\R{{\Bbb R}}
\def\C{{\Bbb C}}
\def\H{{\Bbb H}}

\def\Re{{\rm Re}}
\def\T{{\rm T}}

\def\Mat{\mathrm{Mat}}

\def\diag{{\rm diag}}
\def\cl{{\cal C}\!\ell}

\def\tr{{\rm tr}}

\newcommand{\be}{\begin{equation}}
\newcommand{\ee}{\end{equation}}

\def\SU{{\rm SU}}

\def\U{{\rm U}}
\def\su{{\rm su}}
\def\u{{\rm u}}

\def\dd{{\rm 2x2}}

\newtheorem{theorem}{Theorem}

\begin{document}

\title{Conservative field equations \\ and scalar fields (equations for leptons)}

\maketitle


\author{Nikolay Marchuk\footnote{1) Steklov Mathematical Institute of Russian Academy of Sciences;

2) National Research University Higher School of Economics,

orcid.org/0000-0001-6185-434X}
}

\begin{abstract}
This paper proposes SU(2)-gauge-invariant field equations involving interaction with a scalar field. A connection with the Dirac  equation is discussed.
 \end{abstract}

\medskip

MSC 70S15; 15A66; 15A75


\tableofcontents

\bigskip


The article \cite{DE2} introduces conservative \dd-equations with the unitary group $\U(2)$ as the gauge
 quasysymmetry\footnote{The concept of {\em quasysymmetry} is introduced in the article \cite{DE2}, section 12.} group and proposes applying these equations to describe the dynamics of a non-zero-mass neutrino interacting with the $\SU(2)$ Yang-Mills field (with the weak interaction field). This article builds on the results of the article \cite{DE2}.
 Here we propose SU(2)-gauge-invariant field equations involving interaction with a scalar field. A connection with the Dirac  equation is discussed.

 In this paper (as in the previous papers \cite{DE2},\cite{TMP1},\cite{TMP2}), all physical constants, except for particle masses, are taken to be equal to one.
 Issues related to secondary quantization, renormalization, and the Lagrangian formalism are beyond the scope of this article.


\section{Second-order matrices in Minkowski space}

In the equations presented in this article, the wave functions of particles are described by second-order matrices in Minkowski space.

\medskip

\noindent{\bf Second-order matrices.} Let $\Mat(2,\C)$ denote the algebra of complex $2\times2$ matrices. The matrix algebra $\Mat(2,\C)$ is isomorphic to the biquaternion algebra $\C\otimes\H$ and to the complexified Clifford algebra $\C\otimes\cl_{0,2}$ (see \cite{MarShir}), which is used in some approaches to field theory equations (see \cite{Lanczos:1}, \cite{Mar2018}) on which we rely.

Let $\sigma^0=e$ be the $2\times2$ identity matrix, and let $\sigma^1$, $\sigma^2$, and $\sigma^3$ be the Pauli matrices.
 Let $\pi_+$ be the projection operator from the four-dimensional complex vector space $\Mat(2,\C)$ onto the one-dimensional subspace of complex matrices proportional to the identity matrix $e$, and $\pi_-$ be the projection operator onto the three-dimensional subspace of traceless complex matrices
$$
\pi_+(A) :=(\frac{1}{2}\tr A)e,\quad \pi_-(A) := A - \pi_+(A),\quad \forall A\in \Mat(2,\C).
$$
Any matrix $A\in \Mat(2,\C)$ can be written in the form
 $A=\pi_+(A)+\pi_-(A)=A_+ + A_-$.
Let us introduce the quaternion conjugation operation $A\to\tilde A := \pi_+(A)-\pi_-(A)$ and the superposition of the Hermitian conjugation and the quaternion conjugation $A\to A^* := (\tilde A)^\dagger = \widetilde{(A^\dagger)}$.
For non-singular second-order matrices, let us also introduce the conjugation operation
\begin{equation}
A\to\hat A :=\frac{\det A}{|\det A|}A^*.\label{hat:conj}
\end{equation}
We have
$$
(A B)^\dagger=B^\dagger A^\dagger,\quad \widetilde{A B}=\tilde B \tilde A, \quad (A B)^* = A^* B^*,\quad \widehat{A B}=\hat{A}\hat{B},
\quad \forall A,B\in\Mat(2,\C).
$$
The properties of the introduced conjugation operations are discussed in detail in \cite{DE2}.
\medskip

\noindent{\bf Matrices in Minkowski space.} Let $\R^{1,3}$ be Minkowski space with Cartesian coordinates $x^\mu$, where $\mu = 0, 1, 2, 3$, with partial derivatives $\partial_\mu = \frac{\partial}{\partial x^\mu}$. The metric tensor is given by the Minkowski diagonal matrix $\eta=\diag(1,-1,-1,-1)$.

From here on, we will restrict our analysis to the bounded domain $\Omega\subset\R^{1,3}$. We consider matrices in the domain $\Omega$ as (matrix-valued) functions $\Omega\to\Mat(2,\C)$. We assume that the smoothness of the functions used is sufficient for the validity of the arguments presented\footnote{In this work, it suffices to assume that all functions used are twice continuously differentiable in the domain $\Omega$.}.

 From here on, we will use left (from $(\ell)$) and right (from $(\ell^*)$) \dd-spinors (spinor fields) as specified in \cite{DE2}, as well as tensors (tensor fields) with values in matrices (see \cite{DE2}).


\section{Left- and right-handed conservative equations}

The left- and right-handed conservative \dd-equations were introduced in \cite{DE2} as modifications of the Lanczos quaternion equations (see \cite{Lanczos:1}, \cite{Gsponer}).

In what follows, we will use anti-Hermitian matrices $N\in \u(2)$ in our equations that satisfy an additional condition. We will formulate some properties of such matrices in the form of a theorem.
\begin{theorem}
\begin{enumerate}
\item The condition $N\in \u(2)$ is satisfied if and only if $N^*\in \u(2)$.
\item For $N\in \u(2)$, the condition $N^*N=-e$ is satisfied if and only if $\det N=1$.
\item Let $N$ be a matrix such that $N\in \u(2)$ and $\det N=1$. Let $\acute N = V^{-1}N V$, where $V\in \U(2)$. Then $\acute N\in \u(2)$ and $\det\acute N=1$.
\end{enumerate}
\end{theorem}

Let us write down the general form of matrices $N$ satisfying the conditions $N\in \u(2)$, $\det\,N=1$, and $\tr\,N=i\rho$
\begin{equation}
N = V^{-1}\begin{pmatrix}i\lambda & 0\cr 0 & -\frac{i}{\lambda}\end{pmatrix} V,\label{gformN}
\end{equation}
 where $V\in \U(2)$, $\lambda,\rho\in\R$, $\lambda\neq0$, $\lambda\neq\pm1$, $\lambda-1/\lambda=\rho$.

Equation (67) of \cite{DE2} introduces the {\em left-conservative \dd-equation}
\begin{equation}
\tilde\sigma^\mu\partial_\mu\Phi +  m\hat\Phi N=0,\label{lce0}
\end{equation}
where $\Phi=\Phi(x)\in(\ell)$, $x\in\Omega\subset\R^{1,3}$ and
\begin{equation}
N\in \u(2),\quad \det\,N=1,\quad  \tr\,N=i\rho,\quad \rho\in\R.\label{Nro:cond}
\end{equation}
In addition, we assume that $\det\Phi \neq 0$ throughout the entire region $\Omega$ of Minkowski space in which equation (\ref{lce0}) holds.

Equation (\ref{lce0}) yields a differential conservation law (see \cite{DE2}, equation (60)) in the domain $\Omega$
\begin{equation}
\partial_\mu J^\mu=0, \quad \forall x\in\Omega\label{div:L}
\end{equation}
where\footnote{$\u(2)\T^1$ denotes the set of vector fields (tensors of type $(1,0)$) with values in the Lie algebra $\u(2)$ (see \cite{DE2}).}
$$
J^\mu = J^\mu(x) = i\Phi^\dagger\tilde\sigma^\mu\Phi\in \u(2)\T^1.
$$

Equation (\ref{lce0}) is quasi-invariant under a global transformation (independent of $x\in\Omega$) with matrix $V\in \U(2)$
$$
\Phi\to\acute\Phi=\Phi V,\quad N\to\acute N=V^{-1}N V.
$$

Applying the conjugation operator $*$ to both sides of equation (\ref{lce0}) and letting $\Theta=\Phi^*$, we obtain the {\em right conservative \dd-equation} (see \cite{DE2}, equation (69))
\begin{equation}
\sigma^\mu\partial_\mu\Theta + m\hat\Theta N^*=0\label{rce0}
\end{equation}
 with a \dd-spinor $\Theta=\Theta(x)\in(\ell^*)$, $\det\Theta\neq 0$ and with a matrix $N$ that savisfies (\ref{Nro:cond}). Equation (\ref{rce0}) gives the conservation law
\begin{equation}
\partial_\mu(i\Theta^\dagger\sigma^\mu\Theta) = 0,\quad \forall x\in\Omega.\label{div:R}
\end{equation}


\section{The equation for the neutrino and the equation for the antineutrino (second approximation)}

As a first approximation to the equations for neutrinos and antineutrinos, equations (24) and (28) from \cite{TMP1} were considered. As a second approximation to the equations for neutrinos and antineutrinos, equations (53) and (54) from \cite{DE2} were considered.

In this section, we will review these equations and introduce a refinement: we will substitute the currents $\pi_-(i\Phi^\dagger\tilde\sigma^\nu\Phi)$ and $\pi_-(i\Theta^\dagger\sigma^\nu\Theta)$,
which allow us to restrict the gauge symmetry of the equations to the Lie group $\SU(2)$. This reduction is important because the neutrino has no electric charge and, within the framework of the present project, must possess only the $\SU(2)$ gauge symmetry of weak interactions. In addition, we use a new notation for the matrix $N = i H\in\u(2)$.

Let us substitute the Yang-Mills potential $A_\mu=A_\mu(x)\in \su(2)\T_1$ into equation (\ref{lce0}) and
  combine the resulting left-hand conservative \dd-equation with the Yang-Mills equations, the right-hand side of which includes the (matrix-valued) current
  \begin{equation}
  J^\nu = \pi_-(i\Phi^\dagger\tilde\sigma^\nu\Phi)\in \su(2)\T^1.\label{Jnu:neu}
  \end{equation}
  We obtain a system of equations (the $N$-matrix satisfies the conditions in (\ref{Nro:cond}))
 \begin{eqnarray}
 && \tilde\sigma^\mu(\partial_\mu\Phi + \Phi A_\mu) +  m\hat\Phi N=0,\nonumber\\
 && \partial_\mu A_\nu -\partial_\nu A_\mu - [A_\mu, A_\nu] = F_{\mu\nu},\label{main:U2:0}\\
 && \partial_\mu F^{\mu\nu} - [A_\mu, F^{\mu\nu}] = J^\nu,\nonumber\\
 && \det\,\Phi\neq 0,\nonumber
 \end{eqnarray}
 which is quasi-invariant
 under\footnote{From a formal point of view, the system of equations (\ref{main:U2:0}) is (quasi-)invariant under the gauge transformation (\ref{gauge:AFJ:z}) with the symmetry group $\U(2)$. However, the fact that the right-hand side of the Yang-Mills equations contains a current $J^\nu\in \SU(2)\T^1$ justifies restricting the group of gauge (quasi)-symmetry to the group $\SU(2)$.} the
$\SU(2)$  gauge transformation ($V=V(x)\in \SU(2)$)
\begin{eqnarray}
\Phi &\to& \acute\Phi = \Phi V,\nonumber\\
N &\to& \acute N = V^{-1}N V,\nonumber\\
A_\mu &\to& \acute A_\mu = V^{-1}A_\mu V - V^{-1}\partial_\mu V,\label{gauge:AFJ:z}\\
F_{\mu\nu} &\to& \acute F_{\mu\nu} = V^{-1}F_{\mu\nu} V,\nonumber\\
J^\nu &\to& \acute J^\nu = V^{-1}J^\nu V.\nonumber
\end{eqnarray}

Note that from the transformation $N \to \acute N = V^{-1}N V$ with the matrix $V=V(x)\in \SU(2)$, it follows that the matrix $N$ depends on $x$ and is unitary similar to the constant matrix (\ref{gformN}).

We consider the system of equations (\ref{main:U2:0}) as a system of equations for a (left-handed) neutrino interacting with the $\SU(2)$ Yang-Mills  field ($A_\mu\in \su(2)\T_1$, $F_{\mu\nu}\in \su(2)\T_2$).

It is easy to see that for any matrix $A\in \su(2)$ we have $A^*=A$.

Let us consider the system of equations obtained from the system of equations (\ref{main:U2:0}) by applying the conjugation operation $*$ (as well as the definition $\Theta=\Phi^*\in(\ell^*)$ and the identities $A_\mu^*=A_\mu$, $F_{\mu\nu}^*=F_{\mu\nu}$)
\begin{eqnarray}
 && \sigma^\mu(\partial_\mu\Theta + \Theta A_\mu) + m\hat\Theta N^* = 0,\nonumber\\
&& \partial_\mu  A_\nu -\partial_\nu A_\mu - [A_\mu, A_\nu] = F_{\mu\nu},\label{main:U21:R:z}\\
 && \partial_\mu F^{\mu\nu} - [A_\mu, F^{\mu\nu}] = -\pi_-(i\Theta^\dagger\sigma^\nu\Theta), \nonumber\\
  && \det\,\Theta\neq 0,\quad \det\, N^*=1, \quad \tr\, N^*=-i\rho.\nonumber
 \end{eqnarray}
This as a system of equations for a (right-handed) antineutrino interacting with the Yang-Mills field $(A_\mu, F_{\mu\nu})$ with $\SU(2)$ gauge quasi-symmetry.

We also note that the systems of equations (\ref{main:U2:0}) and (\ref{main:U21:R:z}) contain a scalar real parameter $\rho$. The physical interpretation of this parameter is not yet clear.

The Standard Model employs the Higgs mechanism to generate the masses of fundamental fermions and to introduce massive $Z^0$ and $W^\pm$ bosons. When using equations (\ref{main:U2:0}) and (\ref{main:U21:R:z}) to describe neutrinos and antineutrinos, there is no need to use the Higgs mechanism to introduce neutrino and antineutrino masses, since the mass $m$ is already present in the equations. However, the Higgs mechanism is still necessary to describe the massive $Z^0$ and $W^\pm$ bosons.



\section{The equation for the electron and the equation for the positron (second approximation)}

Let $N = N(x) \in \u(2)$ be an anti-Hermitian matrix (a matrix-valued function) such that
$$
\det N = 1,\quad \tr N = i\rho,\quad \forall x \in \Omega
$$
where $\rho$ is a nonzero real constant. And let
\begin{eqnarray*}
&& \Phi=\Phi(x)\in(\ell),\quad \Theta=\Theta(x)\in(\ell^*),\\ && A_\mu=A_\mu(x)\in u(2)\T_1,\quad F_{\mu\nu}=F_{\mu\nu}(x)\in \u(2)\T_2,\quad \forall x\in\Omega.
\end{eqnarray*}
\medskip

\noindent{\bf An equation for the electron.} Let us consider a system of equation for left and right \dd-spinors
$\Phi\in(\ell)$, $\Theta\in(\ell^*)$ and for Yang-Mills field $(A_\mu, F_{\mu\nu})$
 \begin{eqnarray}
 && \tilde\sigma^\mu(\partial_\mu\Phi + \Phi A_\mu) +  m\hat\Phi N=0,\label{main:U2:1}\\
 && \sigma^\mu(\partial_\mu\Theta + \Theta A_\mu) +  m\hat\Theta N=0,\label{main:U2:2}\\
  && \partial_\mu A_\nu -\partial_\nu A_\mu - [A_\mu, A_\nu] = F_{\mu\nu},\label{main:U2:3}\\
 && \partial_\mu F^{\mu\nu} - [A_\mu, F^{\mu\nu}] = J^\nu,\label{main:U2:4}
  \end{eqnarray}
  where
  \begin{equation}
  J^\nu = i\Phi^\dagger\tilde\sigma^\nu\Phi + \pi_+(i\Theta^\dagger\sigma^\nu\Theta) =
  \pi_-(i\Phi^\dagger\tilde\sigma^\nu\Phi)  + \big(\pi_+(i\Phi^\dagger\tilde\sigma^\nu\Phi) + \pi_+(i\Theta^\dagger\sigma^\nu\Theta)\big).\label{Jnu:electron}
  \end{equation}
    Real constants $m,\rho$ are suppose to be nonzero ($\tr N=i\rho$).

 Note that the left and right components $\Phi$ and $\Theta$ of the electron wave function contribute unequally to the current $J^\nu$.
 Specifically, the term $\pi_-(i\Phi^\dagger\tilde\sigma^\nu\Phi)\in \su(2)\T^1$
 includes only the left \dd-spinor $\Phi$, which corresponds to the $V-A$ theory of weak interactions (see Gell-Mann, Feynman, Marshak, Sudarshan 1957).
Whereas to the second term of the current, $\pi_+(i\Phi^\dagger\tilde\sigma^\nu\Phi) + \pi_+(i\Theta^\dagger\sigma^\nu\Theta)\in \u(1)\T^1$, which is associated with electric current, both components (left and right) $\Phi$ and $\Theta$ of the wave function contribute equally.

 The system of equations (\ref{main:U2:1})-(\ref{Jnu:electron}) is quasi-invariant
under the following $\U(2)$ gauge transformation ($V=V(x)\in \U(2)$):
\begin{eqnarray}
\Phi &\to& \acute\Phi = \Phi V,\nonumber\\
\Theta &\to& \acute\Theta = \Theta V,\nonumber\\
N &\to& \acute N = V^{-1}N V,\nonumber\\
A_\mu &\to& \acute A_\mu = V^{-1}A_\mu V - V^{-1}\partial_\mu V,\label{gauge:AFJ:1}\\
F_{\mu\nu} &\to& \acute F_{\mu\nu} = V^{-1}F_{\mu\nu} V,\nonumber\\
J^\nu &\to& \acute J^\nu = V^{-1}J^\nu V.\nonumber
\end{eqnarray}

Within the framework of the proposed conservative model, we consider the system of equations (\ref{main:U2:1})-(\ref{Jnu:electron}) as a system of equations for the electron interacting with the $\U(2)$ Yang- -Mills  field ($A_\mu\in \u(2)\T_1$, $F_{\mu\nu}\in \u(2)\T_2$).
\medskip

\noindent{\bf On the self-consistency of the system of equations (\ref{main:U2:1})-(\ref{Jnu:electron}) for the electron.} The accumulated knowledge regarding mathematical models in mechanics and physics indicates that the equations of a model must not contain contradictions; that is, they must be self-consistent. The development of this concept, as applied to partial differential equations considered in certain regions of Euclidean or pseudo-Euclidean space, leads to the concept of Adamar correctness for the corresponding (initial) boundary value problems for differential equations. A proof of Adamar correctness involves proving the existence of solutions to the boundary value problem from a suitable functional class, proving the uniqueness of the solution, and proving the continuous dependence of the solution on the boundary (and/or initial) conditions and the right-hand sides of the equations.

Now, for the system of equations (\ref{main:U2:1})-(\ref{Jnu:electron}), let us consider a simpler question regarding the consistency of the consequences of the equations. Namely, from equations (\ref{main:U2:1}) and (\ref{main:U2:2}) we obtain the following consequences (see \cite{DE2}, Theorem 5)
$$
\partial_\mu(i\Phi^\dagger\tilde\sigma^\mu\Phi)-[A_\mu, i\Phi^\dagger\tilde\sigma^\mu\Phi]=0,\quad \partial_\mu(i\Theta^\dagger\sigma^\mu\Theta)-[A_\mu, i\Theta^\dagger\sigma^\mu\Theta]=0.
$$
From these equations, using the projection operators $\pi_+$ and $\pi_-$, we obtain the following equations
\begin{eqnarray}
&& \partial_\mu\pi_+(i\Phi^\dagger\tilde\sigma^\mu\Phi)=0,\quad
\partial_\mu\pi_-(i\Phi^\dagger\tilde\sigma^\mu\Phi)-[A_\mu, \pi_-(i\Phi^\dagger\tilde\sigma^\mu\Phi)]=0,\label{nonabel:cons1} \\
&& \partial_\mu\pi_+(i\Theta^\dagger\sigma^\mu\Theta)=0,\quad
\partial_\mu\pi_-(i\Theta^\dagger\sigma^\mu\Theta)-[A_\mu, \pi_-(i\Theta^\dagger\sigma^\mu\Theta)]=0.\label{nonabel:cons2}
\end{eqnarray}
On the other hand, the Yang-Mills equations (\ref{main:U2:3})-(\ref{main:U2:4}) imply that
\begin{equation}
\partial_\mu J^\mu - [A_\mu, J^\mu] = 0.\label{JJJ}
\end{equation}
Substituting the current
$J^\nu$ from (\ref{Jnu:electron}) into (\ref{JJJ}), we obtain the following equality
$$
(\partial_\mu(i\Phi^\dagger\tilde\sigma^\mu\Phi)-[A_\mu, i\Phi^\dagger\tilde\sigma^\mu\Phi]) +
 \partial_\mu\pi_+(i\Theta^\dagger\sigma^\mu\Theta)=0
 $$
 And this equality follows from the equalities (\ref{nonabel:cons1}) and (\ref{nonabel:cons2}). Which proves the consistency (non-contradictory nature) of the system of equations (\ref{main:U2:1})-(\ref{Jnu:electron}).

\medskip

\noindent{\bf The equation for the positron.} Note that if the matrix $A\in \u(2)$, then $A^*\in \u(2)$ and $(\pi_+(A))^*=\pi_+(A^*)$.
The system of equations for the positron can be obtained from the system of equations (\ref{main:U2:1}-\ref{Jnu:electron}) by applying the conjugation operation $*$ to all equations. We obtain the system of equations
\begin{eqnarray}
&& \sigma^\mu(\partial_\mu\Phi^* + \Phi^* A_\mu^*) +  m\hat\Phi^* N^*=0,\nonumber\\
 && \tilde\sigma^\mu(\partial_\mu\Theta^* + \Theta^* A_\mu^*) + m\hat\Theta^* N^*=0,\nonumber\\
   && \partial_\mu A_\nu^* -\partial_\nu A_\mu^* - [A_\mu^*, A_\nu^*] = F_{\mu\nu}^*,\label{main:U2:anti}\\
 && \partial_\mu (F^{\mu\nu})^* - [A_\mu^*, (F^{\mu\nu})^*] = (J^\nu)^*,\nonumber
  \end{eqnarray}
  where
  \begin{eqnarray}
  (J^\nu)^* &=& -i(\Phi^*)^\dagger\sigma^\nu\Phi^* - \pi_+(i(\Theta^*)^\dagger\tilde\sigma^\nu\Theta^*)\nonumber\\
  &=&
  -\pi_-(i(\Phi^*)^\dagger\sigma^\nu\Phi^*)  - \big(\pi_+(i(\Phi^*)^\dagger\sigma^\nu\Phi^*) + \pi_+(i(\Theta^*)^\dagger\tilde\sigma^\nu\Theta^*)\big).\label{Jnu:electron:anti}
  \end{eqnarray}
   This system of equations is quasi-invariant
under the $\U(2)$ gauge transformation (\ref{gauge:AFJ:1}).

 As already mentioned, the equation for the positron in our conservative model is a separate equation (distinct from the equation for the electron). Therefore, for the quantities appearing in the equation for the positron, one could use different symbols instead of $\Phi^*$, $\Theta^*$, $N^*$, $A_\mu^*$, $F_{\mu\nu}^*$, and $(J^\nu)^*$. To simplify the visual comparison of the system of equations for the positron with the system of equations for the electron, it is proposed in (\ref{main:U2:anti}) to re-designate $\Phi^*\to\Theta$, $\Theta^*\to\Phi$, $N^*\to N$,  $A_\mu^*\to A_\mu$, $F_{\mu\nu}^*\to F_{\mu\nu}$, $(J^\nu)^*\to (J^\nu)$. As a result, we obtain the following equation (system of equations) for the positron
(we swap the first and second equations):
 \begin{eqnarray}
 && \tilde\sigma^\mu(\partial_\mu\Phi + \Phi A_\mu) +  m\hat\Phi N=0,\nonumber\\
 && \sigma^\mu(\partial_\mu\Theta + \Theta A_\mu) + m\hat\Theta N=0,\nonumber\\
   && \partial_\mu A_\nu -\partial_\nu A_\mu - [A_\mu, A_\nu] = F_{\mu\nu},\label{main:U2:5}\\
 && \partial_\mu F^{\mu\nu} - [A_\mu, F^{\mu\nu}] = J^\nu,\nonumber
  \end{eqnarray}
  where
  \begin{equation}
  J^\nu = -i\Theta^\dagger\sigma^\nu\Theta - \pi_+(i\Phi^\dagger\tilde\sigma^\nu\Phi) =
  -\pi_-(i\Theta^\dagger\sigma^\nu\Theta)  - \big(\pi_+(i\Theta^\dagger\sigma^\nu\Theta) + \pi_+(i\Phi^\dagger\tilde\sigma^\nu\Phi)\big).\label{Jnu:electron:3}
  \end{equation}
This system of equations (\ref{main:U2:5})-(\ref{Jnu:electron:3}) differs from the system of equations (\ref{main:U2:1})-(\ref{Jnu:electron}) only in the matrix-valued current $J^\mu\in u(2)\T^1$.

Let us write the Yang-Mills field potential $A_\mu\in \u(2)\T_1$ as a sum
$$
A_\mu = i a_\mu e + \dot A_\mu,\quad i a_\mu \in \u(1)\T_1,\ \dot A_\mu\in \su(2)\T_1
$$
and note that
\begin{equation}
A_\mu^* = -i a_\mu e + \dot A_\mu.\label{Astar}
\end{equation}

\medskip

\noindent{\bf A note on the charge conjugation operation of the Dirac equation.} The standard Dirac equation for the electron describes both electrons and positrons simultaneously (a consistent theory of electrons and positrons is achieved in quantum electrodynamics). This feature can be regarded as a strength of the Dirac equation, which had led to the prediction of the existence of the positron even before its experimental discovery. However, at the same time, this circumstance creates difficulties in interpreting the Dirac equation. Within the framework of a conservative model, particles (leptons) and antiparticles are described by different equations, rather than by a single equation (or system of equations) as is the case with the Dirac equation.

Recall that the charge conjugation operation $\psi\to\psi^c$ (see \cite{Shveber}) transforms the wave function of the Dirac equation
$$
\gamma^\mu(\partial_\mu\psi + i a_\mu\psi) + i m\psi =0
$$
into the wave function of the charge-conjugate equation
$$
\gamma^\mu(\partial_\mu\psi^c - i a_\mu\psi^c) + i m\psi^c =0.
$$
The transition $\psi\to\psi^c$
is interpreted as a transition from the wave function of a particle to the wave function of its antiparticle.
In this case, $i a_\mu\to -i a_\mu$, which indicates an analogy with equation (\ref{Astar}) in the transition $A_\mu\to A_\mu^*$.


\section{Equations for the Yang-Mills field interacting with a scalar field}\label{chap5}

Consider, in the domain $\Omega\subset\R^{1,3}$, a system of equations for matrices (scalar and tensor fields with values in second-order matrices) $N_-\in \su(2)$, $A_\mu, J_\mu\in \su(2)\T_1$, $F_{\mu\nu}\in \su(2)\T_2$
 \begin{eqnarray}
 && \partial_\mu(\partial^\mu N_--[A^\mu, N_-]) - [A_\mu, \partial^\mu N_--[A^\mu, N_-]] + m_0{}^2 N_-=0,\label{YMM1}\\
   && \partial_\mu A_\nu -\partial_\nu A_\mu - [A_\mu, A_\nu] = F_{\mu\nu},\label{YMM2}\\
 && \partial_\mu F^{\mu\nu} - [A_\mu, F^{\mu\nu}] - \\
 && \alpha(\partial^\nu N_- -[A^\nu, N_-]) - \beta[N_-, \partial^\nu N_- -[A^\nu, N_-]]  = J^\nu\label{YMM3}
 \end{eqnarray}
 with parameters $\alpha, \beta, m_0\in\R$.
This system of equations is invariant\footnote{From this point on, we return to the concepts of {\em invariance  and symmetry} of equations, rather than the concepts of {\em quasi-invariance  and quasi-symmetry} of equations, since we will henceforth consider the $N_-$ matrix as the wave function of a scalar particle, rather than as the coefficient matrix of the equation.}
 with respect to the following gauge transformation with the symmetry group $\SU(2)$:
\begin{eqnarray}
N_- &\to& \acute N_- = V^{-1}N_- V,\nonumber\\
A_\mu &\to& \acute A_\mu = V^{-1}A_\mu V - V^{-1}\partial_\mu V,\label{gauge:YMM}\\
F_{\mu\nu} &\to& \acute F_{\mu\nu} = V^{-1}F_{\mu\nu} V,\nonumber\\
J^\nu &\to& \acute J^\nu = V^{-1}J^\nu V,\nonumber
\end{eqnarray}
where $V=V(x)\in \SU(2)$.

Let us move the terms with coefficients $\alpha$ and $\beta$ to the right-hand side of the Yang-Mills equation (\ref{YMM3}) and denote
$$
\check{J}^\nu = \alpha(\partial^\nu N_- -[A^\nu, N_-]) + \beta[N_-, \partial^\nu N_- -[A^\nu, N_-]] + J^\nu.
$$
Let us denote
$$
\dot J^\nu = \beta[N_-, \partial^\nu N_- -[A^\nu, N_-]].
$$
Note that, by virtue of equation (\ref{YMM1}), the following equality holds:
\begin{equation}
\partial_\nu\dot J^\nu - [A^\nu, \dot J^\nu] = 0.\label{Jdot}
\end{equation}
Therefore, as a consequence of equations (\ref{YMM1})-(\ref{YMM3}), we obtain the equality $\partial_\nu \check J^\nu - [A_\nu, \check J^\nu] = 0$, which, taking into account equation (\ref{YMM1}) and equality (\ref{Jdot}), can be written as
\begin {equation}
\partial_\nu J^\nu- [A_\nu, J^\nu]= \alpha m_0{}^2 N_-.\label{JJM}
\end{equation}

In the following sections, the system of equations (\ref{YMM1})-(\ref{YMM3}) with the condition (\ref{JJM}) is used to further develop the conservative neutrino model under discussion.

Note that the system of equations (\ref{YMM1})-(\ref{YMM3}) generalizes the Yang-Mills-Higgs system considered in a number of works (\cite{Taubes1}, \cite{Kapitanskii1987}, etc.) and containing the term $\dot J^\nu$.


\section{The equation for the neutrino and the equation for the antineutrino (third approximation)}

Let us consider the system of equations
 \begin{eqnarray}
 && \tilde\sigma^\mu(\partial_\mu\Phi + \Phi A_\mu) + m\lambda_1\hat\Phi N=0,\label{third:1}\\
 && \partial_\mu A_\nu -\partial_\nu A_\mu - [A_\mu, A_\nu] = F_{\mu\nu},\label{third:2}\\
 && \partial_\mu F^{\mu\nu} - [A_\mu, F^{\mu\nu}] =\label{third:3}\\
 && \alpha(\partial^\nu N_- -[A^\nu, N_-])+\beta[N_-, \partial^\nu N_- -[A^\nu, N_-]] +
 \pi_-(i\Phi^\dagger\tilde\sigma^\nu\Phi),\nonumber\\
 && \partial_\mu(\partial^\mu N_- -[A^\mu, N_-]) - [A_\mu, \partial^\mu N_- -[A^\mu, N_-]] + m_0{}^2 N_-=0,\label{third:4}\\
 && \det\,N=1,\label{third:5}
 \end{eqnarray}
where
\begin{eqnarray*}
&& \Phi\in(\ell),\quad A_\mu\in \su(2)\T_1,\quad F_{\mu\nu}\in \su(2)\T_2,\quad N\in \u(2),\\
&&\alpha, \beta, m, m_0\in\R,\quad \lambda_1\in\C,\quad |\lambda_1|=1.
\end{eqnarray*}
We assume that $m$ and $m_0$ are nonzero constants, while the remaining quantities $\Phi$, $A_\mu$, $F_{\mu\nu}$, $N$, $\alpha$, $\beta$, and $\lambda_1$ depend on $x \in \Omega \subset \mathbb{R}^{1,3}$. We also assume that
$$
\det\Phi\neq 0,\quad \alpha\neq 0,\quad \Re\lambda_1\neq 0,\quad \forall x\in\Omega.
$$
Let us introduce a notation for the vector with values in the Lie algebra $\su(2)$, which appears on the right-hand side of the Yang-Mills equation (\ref{third:3})
\begin{equation}
J^\nu := \alpha(\partial^\nu N_- -[A^\nu, N_-])+\beta[N_-, \partial^\nu N_- -[A^\nu, N_-]] +\pi_-(i\Phi^\dagger\tilde\sigma^\nu\Phi)\in \su(2)\T^1.\label{Jnu}
\end{equation}
The system of equations (\ref{third:1})-(\ref{third:5}) is invariant under the $\SU(2)$ gauge transformation (\ref{gauge:AFJ:z}),
and the values of $m, m_0, \alpha, \beta, \lambda_1$
remain unchanged under this gauge transformation.

We consider the system of equations (\ref{third:1}) (\ref{third:5}) as a system of equations (in the third approximation) for a neutrino interacting with the $\SU(2)$ Yang-Mills gauge field ($A_\mu\in \su(2)\T_1$, $F_{\mu\nu}\in \su(2)\T_2$).

Let us consider some consequences of the system of equations (\ref{third:1})-(\ref{third:5}) and find conditions on the values of $\alpha$ and $\lambda_1$ under which the consequences of the equations are consistent\footnote{The consistency of the equations can be achieved by applying the Lagrangian formalism. This approach to field equations is beyond the scope of this article.}.

First,
multiply the left-hand side of equation (\ref{third:1}) by $\Phi^\dagger$ and add the result to the Hermitian conjugate of the expression. We get (after multiplying by $i$)
\begin{equation}
\partial_\mu(i\Phi^\dagger\tilde\sigma^\mu\Phi) -[A_\mu, i\Phi^\dagger\tilde\sigma^\mu\Phi]= 2m\epsilon_1|\det\Phi|N,\label{cons:1}
\end{equation}
where we use the decomposition of the complex function $\lambda_1 = \rho_1 + i\epsilon_1$ into its real and imaginary parts ($\rho_1, \epsilon_1 : \Omega \to \mathbb{R}$).

Let us apply the projection operators $\pi_+$ and $\pi_-$ to both sides of the equality (\ref{cons:1}). We obtain
\begin{eqnarray}
&& \partial_\mu\pi_-(i\Phi^\dagger\tilde\sigma^\mu\Phi) -[A_\mu, \pi_-(i\Phi^\dagger\tilde\sigma^\mu\Phi)]= 2m\epsilon_1|\det\Phi|N_-,\label{cons:2}\\
&& \partial_\mu\pi_+(i\Phi^\dagger\tilde\sigma^\mu\Phi)= 2m\epsilon_1|\det\Phi|N_+.\label{cons:3}
\end{eqnarray}

Second, let us derive a consequence from the Yang-Mills equations (\ref{third:2})-(\ref{third:3})
\begin{eqnarray}
0 &=& \partial_\nu J^\nu - [A_\nu, J^\nu]\label{cons:4} \\
&=& (\partial_\nu\alpha)(\partial^\nu N_- - [A^\nu, N_-])
+ (\partial_\nu\beta)[N_-, \partial^\nu N_- -[A^\nu, N_-]]\nonumber\\
 &&+ \alpha(\partial_\nu(\partial^\nu N_- - [A^\nu, N_-]) - [A_\nu, \partial^\nu N_- - [A^\nu, N_-]])\nonumber\\
  &&+ (\partial_\nu\pi_-(i\Phi^\dagger\tilde\sigma^\mu\Phi) - [A_\nu, \pi_-(i\Phi^\dagger\tilde\sigma^\mu\Phi)]).\nonumber
\end{eqnarray}
Equation (\ref{third:4}) gives
\begin{eqnarray}
&& \partial_\mu(\partial^\mu N_--[A^\mu, N_-]) - [A_\mu, \partial^\mu N_--[A^\mu, N_-]] = - m_0{}^2 N_-.\label{third:42}
\end{eqnarray}

From equation (\ref{cons:4}), using relations (\ref{third:42}) and (\ref{cons:2}), we obtain the following condition for real-valued functions $\alpha$, $\beta$, and $\epsilon_1$:
\begin{eqnarray}
&& (\partial_\nu\alpha)(\partial^\nu N_- - [A^\nu, N_-])+ (\partial_\nu\beta)[N_-, \partial^\nu N_- -[A^\nu, N_-]] \nonumber\\
&&-\alpha m_0{}^2 N_- + 2m\epsilon_1|\det\Phi|N_-=0,\label{main:cond}
\end{eqnarray}
providing consistency for the system of equations (\ref{third:1})-(\ref{third:5}).

Let us give a particular solution to equation (\ref{main:cond}) obtained under the assumption that $\alpha$ and $\beta$ are real constants. In this case, from (\ref{main:cond}) we obtain
$$
\alpha = \frac{2m}{m_0^2}\epsilon_1|\det\Phi|.
$$
By assumption, $\alpha$ does not depend on $x\in\Omega$, while $|\det\Phi|$ depends on $x$. Therefore, we must take
$$
\epsilon_1 = \frac{\epsilon}{|\det\Phi|},\quad \alpha = \frac{2m\epsilon}{m_0^2},
$$
where $\epsilon$ is some nonzero real constant.
Since $\lambda_1=\rho_1 + i\epsilon_1$ and $|\lambda_1|=1$, we have $\rho_1^2+\epsilon_1^2=1$ and
$$
\rho_1 = \pm\frac{1}{|\det\Phi|}\sqrt{|\det\Phi|^2-\epsilon^2}.
$$
This leads to an additional constraint on the matrix-valued function $\Phi$
\begin{equation}
|\det\Phi| > |\epsilon|,\quad\forall x\in\Omega.\label{detPhi:eps}
\end{equation}

Thus, it has been proven that if we take the functions $\lambda_1 : \Omega\to\C$ and $\alpha : \Omega\to\R$ to depend on a nonzero real constant $\epsilon$ according to the formulas (in the assumption (\ref{detPhi:eps}))
\begin{equation}
\lambda_1 = \frac{1}{|\det\Phi|}\big(\pm\sqrt{|\det\Phi|^2 - \epsilon^2} + i\epsilon\big), \quad
\alpha = \frac{2m\epsilon}{m_0^2},\label{lam1:alpha:cond}
\end{equation}
then the consequences (\ref{cons:1}) and (\ref{cons:4}) of the system of equations (\ref{third:1})-(\ref{third:5}) will be consistent (non-contradictory).
In this case, the system of equations (\ref{third:1})-(\ref{third:5}) can be regarded as a system of equations describing the dynamics of a neutrino with a nonzero mass $m$.

Equation (\ref{third:1}) involves the matrix-valued function $N : \Omega \to \u(2)$, where $N=\pi_+(N) + \pi_-(N)=N_+ + N_-$. Equations (\ref{third:4}) and (\ref{third:5}) are considered as equations for the function $N$. In particular, the equation $\det\,N=1$ allows us to express $N_+$ in terms of
$N_-$.

  Indeed, we have $\tilde N = N_+ - N_-$, $N\tilde N = (\det\,N)e$. Therefore, when $\det\,N = 1$, we see that
$$
N\tilde N = (N_+ + N_-)(N_+ - N_-) = N_+^2 - N_-^2 = e, \quad N_+ = \pm\sqrt{e + N_-^2}.
$$
If we express $N\in\u(2)$ in terms of the Pauli basis, we obtain
$$
N=i n_k\sigma^k,\quad n_l\in\mathbb{R},\quad l=0,1,2,3
$$
and
$$
N_+ = i n_0 e,\quad N_- = i n_1\sigma^1 +  i n_2\sigma^2 +  i n_3\sigma^3.
$$
Since
$$
\det\,N = - n_0^2 + n_1^2 + n_2^2 + n_3^2,
$$
the condition $\det\,N=1$ yields
$$
n_0 = \pm\sqrt{n_1{}^2 + n_2{}^2 + n_3{}^2 -1},\quad n_1{}^2 + n_2{}^2 + n_3{}^2\geq 1.
$$
Equation (\ref{third:4}) is a generalization of the Klein-Gordon equation to scalar functions $N_- : \Omega\to \su(2)$ (with the gauge symmetry group $\SU(2)$). Therefore, we can assume that the quanta of the scalar field $N_-$ are scalar particles with a nonzero mass $m_0$.


\medskip

\noindent{\bf The equation for the antineutrino (third approximation)}.
To obtain a system of equations for the (right-handed) antineutrino, we apply the operator $*$ to all equations
(\ref{third:1})-(\ref{third:5}). Let $\Theta=\Phi^*$, let $N^*$ denote $N$, and use the identities $A_\mu^*=A_\mu$ and $F_{\mu\nu}^*=F_{\mu\nu}$, which hold for matrices in the Lie algebra $su(2)$. As a result, we obtain a system of equations (in the third approximation)
for the antineutrino
\begin{eqnarray}
 && \sigma^\mu(\partial_\mu\Theta + \Theta A_\mu) + m\bar\lambda_1\hat\Theta N=0,\label{third:1a}\\
 && \partial_\mu A_\nu -\partial_\nu A_\mu - [A_\mu, A_\nu] = F_{\mu\nu},\label{third:2a}\\
 && \partial_\mu F^{\mu\nu} - [A_\mu, F^{\mu\nu}] =\label{third:3a}\\
 && \alpha(\partial^\nu N_- -[A^\nu, N_-]) + \beta[N_-, \partial^\nu N_- -[A^\nu, N_-]]-
 \pi_-(i\Theta^\dagger\sigma^\nu\Theta),\nonumber\\
 && \partial_\mu(\partial^\mu N_- -[A^\mu, N_-]) - [A_\mu, \partial^\mu N_- -[A^\mu, N_-]] + m_0{}^2 N_-=0,\label{third:4a}\\
 && \det\,N=1,\quad |\lambda_1|=1,\label{third:5a}
 \end{eqnarray}
where
\begin{eqnarray*}
&& \Theta\in(\ell^*),\quad A_\mu\in \su(2)\T_1,\quad F_{\mu\nu}\in \su(2)\T_2,\quad N\in \u(2),\\
&&\alpha, \beta, m, m_0\in\R,\quad \lambda_1\in\C
\end{eqnarray*}
and
\begin{equation}
\bar\lambda_1 = \frac{1}{|\det\Theta|}\big(\pm\sqrt{|\det\Theta|^2 - \epsilon^2} - i\epsilon\big), \quad
\alpha = \frac{2m\epsilon}{m_0{}^2},\label{lam1:alpha:cond0}
\end{equation}
$\epsilon$ is a nonzero real constant satisfying the condition $|\det\Theta| > |\epsilon|$ for all $x \in \Omega$.

Note that from the equality $\Theta=\Phi^*$ it follows that $|\det\Theta|=|\det\Phi|$.


\section{The equation for the electron and the equation for the positron (third approximation)}\label{chap8}

 Consider the system of equations
 \begin{eqnarray}
 && \tilde\sigma^\mu(\partial_\mu\Phi + \Phi A_\mu) +  m\lambda_1\hat\Phi N=0,\label{third:1y}\\
 && \sigma^\mu(\partial_\mu\Theta + \Theta A_\mu) +  m\lambda_2\hat\Theta N=0,\label{third:1xy}\\
 && \partial_\mu A_\nu -\partial_\nu A_\mu - [A_\mu, A_\nu] = F_{\mu\nu},\label{third:2y}\\
 && \partial_\mu F^{\mu\nu} - [A_\mu, F^{\mu\nu}] = J^\nu,\label{third:3y}\\
 && \partial_\mu(\partial^\mu N_- -[A^\mu, N_- ]) - [A_\mu, \partial^\mu N_- -[A^\mu, N_- ]] + m_0{}^2 N_- =0,\label{third:4y}\\
 && \det\,N=1,\quad |\lambda_1|=|\lambda_2|=1,\label{third:5y}
 \end{eqnarray}
where
\begin{eqnarray}
J^\nu &=& \alpha(\partial^\nu N_-  -[A^\nu, N_- ])  + \beta[N_-, \partial^\nu N_- -[A^\nu, N_-]]\nonumber\\
 &&+ (i\Phi^\dagger\tilde\sigma^\nu\Phi) +\pi_+(i\Theta^\dagger\sigma^\nu\Theta)\in u(2)\T^1\label{Jnu:y}
\end{eqnarray}
and
\begin{eqnarray*}
&& \Phi\in(\ell),\quad \Theta\in(\ell^*),\quad A_\mu\in \u(2)\T_1,\quad F_{\mu\nu}\in \u(2)\T_2,\quad N\in \u(2),\\
&&\alpha,\beta, m, m_0\in\R,\quad \lambda_1,\lambda_2\in\C.
\end{eqnarray*}
Let $m$ and $m_0$ be nonzero  real constants, and let the remaining quantities $\Phi$, $\Theta$, $A_\mu$, $F_{\mu\nu}$, $N$, $\alpha$, $\lambda_1$, and $\lambda_2$ depend on $x \in \Omega \subset \mathbb{R}^{1,3}$. We also assume that
$$
\det\Phi \neq 0, \quad \det\Theta \neq 0, \quad \alpha \neq 0, \quad \Re\lambda_1 \neq 0, \quad \Re\lambda_2 \neq 0, \quad \forall x \in \Omega.
$$

The system of equations (\ref{third:1y})-(\ref{Jnu:y}) is invariant under the following $\U(2)$ gauge transformation ($V=V(x)\in \U(2)$):
\begin{eqnarray}
\Phi &\to& \acute\Phi = \Phi V,\nonumber\\
\Theta &\to& \acute\Theta = \Theta V,\nonumber\\
A_\mu &\to& \acute A_\mu = V^{-1}A_\mu V - V^{-1}\partial_\mu V,\label{gauge:AFJ:0}\\
F_{\mu\nu} &\to& \acute F_{\mu\nu} = V^{-1}F_{\mu\nu} V,\nonumber\\
J^\nu &\to& \acute J^\nu = V^{-1}J^\nu V.\nonumber\\
N &\to& \acute N = V^{-1}N V.\nonumber
\end{eqnarray}
Moreover, the values of $m, m_0, \alpha, \lambda_1, \lambda_2$
remain unchanged under this gauge transformation.

Within the framework of the proposed conservative lepton model, we consider the system of equations (\ref{third:1y})-(\ref{Jnu:y}) as a system of equations for an electron interacting with the $\U(2)$ Yang- -Mills field ($A_\mu\in \u(2)\T_1$, $F_{\mu\nu}\in \u(2)\T_2$).

Let us consider some consequences of the system of equations (\ref{third:1y})-(\ref{Jnu:y}) and determine the conditions on the values of $\alpha, \lambda_1, \lambda_2$ under which the consequences of the equations are not contradictory.

First, from equations (\ref{third:1y}) and (\ref{third:1xy}) we obtain
\begin{eqnarray}
\partial_\mu(i\Phi^\dagger\tilde\sigma^\mu\Phi) -[A_\mu, i\Phi^\dagger\tilde\sigma^\mu\Phi] &=& 2m\epsilon_1|\det\Phi|N,\label{cons:1y}\\
\partial_\mu(i\Theta^\dagger\sigma^\mu\Theta) -[A_\mu, i\Theta^\dagger\sigma^\mu\Theta] &=& 2m\epsilon_2|\det\Theta|N,\label{cons:1xy}
\end{eqnarray}
where we use the decomposition of complex functions $\lambda_k = \rho_k + i\epsilon_k$, $k = 1, 2$ into real and imaginary parts ($\rho_k, \epsilon_k : \Omega \to \mathbb{R}$). Let us act on the equalities (\ref{cons:1y}), (\ref{cons:1xy}) with the projection operators $\pi_+,\pi_-$. We obtain
\begin{eqnarray}
&& \partial_\mu\pi_-(i\Phi^\dagger\tilde\sigma^\mu\Phi) -[A_\mu, \pi_-(i\Phi^\dagger\tilde\sigma^\mu\Phi)]= 2m\epsilon_1|\det\Phi|N_-,\label{cons:2y}\\
&& \partial_\mu\pi_+(i\Phi^\dagger\tilde\sigma^\mu\Phi)= 2m\epsilon_1|\det\Phi|N_+,\label{cons:3y}\\
&& \partial_\mu\pi_-(i\Theta^\dagger\sigma^\mu\Theta) -[A_\mu, \pi_-(i\Theta^\dagger\sigma^\mu\Theta)]= 2m\epsilon_2|\det\Theta|N_-,\label{cons:2xy}\\
&& \partial_\mu\pi_+(i\Theta^\dagger\sigma^\mu\Theta)= 2m\epsilon_2|\det\Theta|N_+.\label{cons:3xy}
\end{eqnarray}

Second, let us derive a consequence from the Yang-Mills equations (\ref{third:2y})-(\ref{third:3y}) (taking into account the equality (\ref{Jdot}))
\begin{eqnarray}
0 &=& \partial_\nu J^\nu - [A_\nu, J^\nu]\label{cons:4y} \\
&=& (\partial_\nu\alpha)(\partial^\nu N_- - [A^\nu, N_-]) + (\partial_\nu\beta)[N_-, \partial^\nu N_- -[A^\nu, N_-]]\nonumber\\
&&+  \alpha(\partial_\nu(\partial^\nu N_- - [A^\nu, N_-]) - [A_\nu, \partial^\nu N_- - [A^\nu, N_-]])\nonumber\\
  &&+ (\partial_\nu(i\Phi^\dagger\tilde\sigma^\mu\Phi) - [A_\nu, i\Phi^\dagger\tilde\sigma^\mu\Phi]) + \partial_\nu\pi_+(i\Theta^\dagger\sigma^\nu\Theta).\nonumber
\end{eqnarray}
From equation (\ref{cons:4y}), using relations (\ref{third:4y}), (\ref{cons:1y}), and (\ref{cons:3xy}), we obtain the following condition for the real functions $\alpha,\beta,\epsilon_1,\epsilon_2$:
\begin{eqnarray}
&& (\partial_\nu\alpha)(\partial^\nu N_- - [A^\nu, N_-])+ (\partial_\nu\beta)[N_-, \partial^\nu N_- -[A^\nu, N_-]] \nonumber\\
&&-\alpha m_0{}^2 N_- + 2m\epsilon_1|\det\Phi|N + 2m\epsilon_2|\det\Theta|N_+=0,\label{main:cond:y}
\end{eqnarray}
which ensures the consistency of the system of equations  (\ref{third:1y})-(\ref{Jnu:y}).

Let us specify a particular solution to equation (\ref{main:cond:y}), obtained under the assumption that $\alpha$ and $\beta$ are  real constants. In this case
\begin{equation}
 -\alpha m_0{}^2 N_- + 2m\epsilon_1|\det\Phi|N + 2m\epsilon_2|\det\Theta|N_+=0,\label{main:cond:z}
\end{equation}
Applying the projection operators $\pi_+$ and $\pi_-$ to this equality, we obtain
\begin{eqnarray}
&& 2m\epsilon_1|\det\Phi|N_+ + 2m\epsilon_2|\det\Theta|N_+ = 0,\label{const:cond:1}\\
&& -\alpha m_0{}^2 N_- + 2m\epsilon_1|\det\Phi|N_- = 0.\label{const:cond:2}
\end{eqnarray}
Since $N_+\neq 0$ and $N_-\neq 0$, from equations (\ref{const:cond:1}) and (\ref{const:cond:2}) we obtain
$$
\epsilon_1=\frac{\epsilon}{|\det\Phi|},\quad
\epsilon_2=-\frac{\epsilon}{|\det\Theta|},\quad
\alpha = \frac{2m\epsilon}{m_0^2},
$$
where $\epsilon$ is some nonzero real constant.
Since
$$
\lambda_k=\rho_k + i\epsilon_k,\quad |\lambda_k|=1,\quad\rho_k{}^2+\epsilon_k{}^2=1,\quad k=1,2,
$$
then
$$
\rho_1 = \pm\frac{1}{|\det\Phi|}\sqrt{|\det\Phi|^2-\epsilon^2},\quad
\rho_2= \pm\frac{1}{|\det\Theta|}\sqrt{|\det\Theta|^2-\epsilon^2}.
$$
In this case, additional constraints arise on the matrix-valued functions $\Phi$ and $\Theta$ and the real parameter $\epsilon$
\begin{equation}
|\det\Phi| > |\epsilon|,\quad |\det\Theta| > |\epsilon|,\quad\forall x\in\Omega.\label{detPhi:eps:y}
\end{equation}

Thus, it has been proven that if we take the functions $\lambda_{1,2} : \Omega\to\C$ and $\alpha : \Omega\to\R$ to depend on a nonzero real constant $\epsilon$ and on $|\det\Phi|$ and $|\det\Theta|$ according to the formulas (under the assumption (\ref{detPhi:eps:y}))
\begin{eqnarray}
&& \lambda_1 = \frac{1}{|\det\Phi|}(\pm\sqrt{|\det\Phi|^2-\epsilon^2} + i\epsilon),\nonumber\\
&& \lambda_2 = \frac{1}{|\det\Theta|}(\pm\sqrt{|\det\Theta|^2-\epsilon^2} - i\epsilon),\label{lam1:alpha:cond}\\
&& \alpha = \frac{2m\epsilon}{m_0^2},\nonumber
\end{eqnarray}
then the consequences (\ref{cons:1y}), (\ref{cons:1xy}) and (\ref{cons:4y}) of the system of equations (\ref{third:1y})-(\ref{Jnu:y}) will be consistent (non-contradictory), and the system of equations (\ref{third:1y})-(\ref{Jnu:y}) can be regarded as a system of equations describing the dynamics of an electron with mass $m$. The signs $\pm$ in the formulas (\ref{lam1:alpha:cond}) mean that all four possible choices of plus and minus signs are valid.

\medskip

\noindent{\bf Equations for the positron (third approximation).} To obtain the equations for the positron, we apply the conjugation operator $*$ to all equations (\ref{third:1y})-(\ref{Jnu:y}), swap the first and second equations, and redefine the variables $\Phi^*\to\Theta$, $\Theta^*\to\Phi$, $N^*\to N$, $A_\mu^*\to A_\mu$, $F_{\mu\nu}^*\to F_{\mu\nu}$, $J_\nu^*\to J_\nu$. We obtain a system of equations
 \begin{eqnarray}
 && \tilde\sigma^\mu(\partial_\mu\Phi + \Phi A_\mu) +  m\bar\lambda_2\hat\Phi N=0,\label{third:1y:anti}\\
 && \sigma^\mu(\partial_\mu\Theta + \Theta A_\mu) +  m\bar\lambda_1\hat\Theta N=0,\label{third:1xy:anti}\\
 && \partial_\mu A_\nu -\partial_\nu A_\mu - [A_\mu, A_\nu] = F_{\mu\nu},\label{third:2y:anti}\\
 && \partial_\mu F^{\mu\nu} - [A_\mu, F^{\mu\nu}] = J^\nu,\label{third:3y:anti}\\
 && \partial_\mu(\partial^\mu N_- -[A^\mu, N_-]) - [A_\mu, \partial^\mu N_- -[A^\mu, N_-]] + m_0{}^2 N_-=0,\label{third:4y:anti}\\
 && \det\,N=1,\quad |\lambda_1|=|\lambda_2|=1,\label{third:5y:anti}
 \end{eqnarray}
where
\begin{equation}
J^\nu = \alpha(\partial^\nu N_- -[A^\nu, N_-])+\beta[N_-, \partial^\nu N_- -[A^\nu, N_-]] -i\Theta^\dagger\sigma^\nu\Theta - \pi_+(i\Phi^\dagger\tilde\sigma^\nu\Phi) \label{Jnu:y:anti}
\end{equation}
and
\begin{eqnarray*}
&& \Phi\in(\ell),\quad \Theta\in(\ell^*),\quad A_\mu,J_\mu\in \u(2)\T_1,\quad F_{\mu\nu}\in \u(2)\T_2,\quad N\in \u(2),\\
&& \alpha, \beta, m, m_0\in\R.
\end{eqnarray*}
In this case, the quantities $\lambda_1,\lambda_2,\alpha$ depend on the real constants $m$, $m_0$, $\epsilon$ and on $|\det\Phi|$, $|\det\Theta|$ according to the formulas (\ref{lam1:alpha:cond})
and the conditions (\ref{detPhi:eps:y}) are satisfied.

The system of equations (\ref{third:1y:anti})-(\ref{Jnu:y:anti}) is invariant under the $\U(2)$ gauge transformation (\ref {gauge:AFJ:0}) and, within the scope of our project, is considered as a system of equations for the positron (as well as for the antimuon and the antitau lepton).


\section{Conclusions.}
In the paper we postulated the $\SU(2)$-gauge-invariant \dd-equations (\ref{third:1})-(\ref{third:5}), which we interpret as equations for neutrinos with nonzero mass, and equations (\ref{third:1a})-(\ref{third:5a}) as equations for antineutrinos.

Also we postulated the $\U(2)$-gauge-invariant \dd-equations (\ref{third:1})-(\ref{third:5}), which we interpret as equations for the electron and equations (\ref{third:1a})-(\ref{third:5a}) as equations for the positron.

\medskip
\noindent{\bf Acknowledgments.}
The author is grateful to the staff of the Department of Mathematical Physics of Steklov Mathematical Institute of the Russian Academy of Sciences as well as to the staff of the Laboratory of
Geometric Algebra and Applications of National Research University Higher School of Economics
for constructive discussions of the results of the work.
\medskip

\noindent{\bf Funding.} The article was prepared within the framework of the project ``Mirror
Laboratories'' HSE University ``Quaternions, geometric algebras and applications''.




\end{document}